\documentclass[aip,preprint]{revtex4-1}
\usepackage{graphicx}
\usepackage{caption}
\usepackage{subcaption}
\usepackage{dcolumn}
\usepackage{bm}
\usepackage{ulem}

\begin{document}

\preprint{AIP/123-QED}

\title{Spreading of magnetic reconnection by electron scale dispersive waves }

\author{Neeraj Jain}
\affiliation{ Max Planck Institute for Solar System Research,
Justus-von-Liebig-Weg 3, 37077, G\"ottingen, Germany.}
\author{J\"org B\"uchner}%
\affiliation{ Max Planck Institute for Solar System Research,
Justus-von-Liebig-Weg 3, 37077, G\"ottingen, Germany.}

\date{\today}

\begin{abstract}
We show  that on electron scales a patch of the localized magnetic reconnection spreads bi-directionally  in a wave like fashion when an external guide-magnetic field in the direction of the electron current is present. The spreading is caused by the propagation of the flow induced and whistler wave modes away from the localized patch. For small guide fields, the spreading is asymmetric being faster in the direction of the electron flow. On increasing the guide field, the spreading becomes increasingly symmetric due to the dominance of the whistler group speed in determining the speed of the spreading. The wave-like spreading of reconnection causes the alternate formation of X- and O-points in the reconnection 
planes separated by half the wavelength of the reconnection wave.  
\end{abstract}

\pacs{Valid PACS appear here}
\maketitle

\section{\label{sec:introduction}Introduction}
Magnetic reconnection is considered to be the cause of the release of magnetic energy
in solar flares, sub-storms in Earth's magnetosphere, sawtooth crashes in tokamaks and
many astrophysical systems, e.g., accretion disk. The release of magnetic energy is enabled by dissipation in self consistently formed current sheets allowing topological changes of magnetic field lines. In collisionless situations, e.g., solar flares and Earth’s magnetosphere, dissipation is weak and magnetic reconnection is very slow until the thickness of the current sheet is of the order of microscopic scales, such as, electron and ion inertial lengths where an effective dissipation is provided by micro-physical plasma processes. At these scales, electron and ion dynamics decouple resulting in a two scale structure (along its thickness) of the current sheet, viz., an electron current sheet with thickness of the order of an electron inertial length, $d_e=c/\omega_{pe}$ embedded inside an ion current sheet with thickness of the order of an ion inertial length, $d_i=c/\omega_{pi}$. Reconnection of field lines takes place inside
the so formed electron current sheets and couples to ion, and then further to very large magnetohydrodynamic scales.

In many plasmas of interest, magnetic reconnection first takes place in a localized region of space and then spreads away from the localized region. For example, in solar observations of two ribbon flares, flare brightening indicative of reconnection has been observed to spread bidirectionally along the polarity inversion line \cite{qiu2009,qiu2010,cheng2012}. Laboratory experiments in Versatile Toroidal Facility (VTF) with a strong guide field also show bidirectional spreading of localized reconnection along the guide field \cite{katz2010} . On the other hand unidirectional spreading of reconnection in the
direction of electron drift velocity was observed in Magnetic Reconnection Experiments
(MRX) without guide field \cite{dorfman2013}. In VINETA-II device, new experiments with varying strength of guide field are planned to study the spreading of reconnection along the guide field \cite{bohlin2014,stechow2015,stechow_private_comm}.

Present understanding of the spreading of localized reconnection associates it with either a wave motion associated with reconnection or the motion of current carriers. In 3-D particle-in-cell (PIC) simulations an ion scale structure of connected 3-D nulls and a reconnection wave due to drift sausage instability driven by the current sheet were observed \cite{buechner1996c,buechner1999b,wiegelmann2000}. As a result, reconnection couples to the drift sausage instability and propagates with it in the direction of the current flow. Hall-MHD simulations of reconnection show that reconnection initialized in a localized region of space propagates as a wave called 'reconnection wave' in the direction of electron drift (opposite to the direction of current) with electron drift speed \cite{huba2002,huba2003}. Such reconnection waves are expected when dominant current carriers are electrons \cite{shay2003,karimabadi2004}. In hybrid simulations, in which ions are the dominant current carriers and electrons just neutralizing background, reconnection was found to spread due to the ion motion rather than by a reconnection wave \cite{karimabadi2004}. When both electrons and ions carry currents, a reconnection X-line can expand bi-directionally since electrons
and ions move in opposite directions \cite{lapenta2006}. A parametric study using three dimensional Hall-MHD simulations showed that both electrons and ions can contribute to the spreading of reconnection depending upon their share of the current \cite{nakamura2012}. All these previous studies were carried out for zero guide field. In case of a finite guide field in the current direction,  Alfven waves can propagate along the guide field and contribute to the bi-directional spreading of localized reconnection perturbation. The mechanism of the spreading changes from a current-carrier-dominated to Alfven-waves-dominated after reaching a critical guide field above which Alfven waves propagate faster than the current carriers \cite{shepherd2012}.

As far as the direction of the spreading is concerned, the experimental results are in good agreement with the theoretical and numerical studies. However, the speed of the spreading is not always matched well. In the MRX experiment \cite{dorfman2013} , reconnection spreads in the direction of the electron flow but with a speed much less than the peak electron drift speed. The latter is the theoretically predicted \cite{huba2002} speed of spreading as electrons are the dominant current carrier and the guide field is negligibly small. Three-dimensional electron-magnetohydrodynamic (EMHD) simulations of an electron current sheet with zero guide field showed that, on electron scales, the speed of spreading can be between zero and the peak electron flow speed depending on the wave number of reconnection perturbation \cite{jain2013} . Such wave number dependence of the speed of spreading was
not identified in any of the earlier studies.

In this paper, we extend our earlier studies \cite{jain2013} on the spreading of localized reconnection at electron scales to the case of finite guide magnetic field in the current direction using an electron-magnetohydrodynamic (EMHD) model. Three dimensional EMHD simulations coupled with linear eigen mode analysis are performed for different strength of the guide field.

In the next section we discuss EMHD approach and the simulation setup. In section \ref{sec:spreading}, we discuss the 3-D simulation results on the spreading of reconnection. In this section, an understanding of the simulation results based on the linear eigen value analysis and local dispersion relation of EMHD will be presented. Finally we summarize our findings in section \ref{sec:conclusion}.

\section{\label{sec:emhd}Electron-MHD approach and simulation setup}

The electron-magnetohydrodynamic (EMHD) approach considers electrons as a dynamically evolving fluid 
and a stationary background of ions. The EMHD approach is valid for spatial scales smaller than the ion inertial length ($d_i$) and time scales smaller than $\omega_{ci}^{-1}$, the inverse ion cyclotron frequency. In EMHD, the electron dynamics is described by electron momentum equation coupled with Maxwell’s equations. An evolution equation for the magnetic field  can be obtained by eliminating the electric field from the electron momentum equation using
Faraday’s law \cite{kingsep90}.
\begin{eqnarray}
\frac{\partial}{\partial t}(\mathbf{B}-d_e^2\nabla^2\mathbf{B})&=&\nabla \times
[\mathbf{v}_e\times (\mathbf{B}-d_e^2\nabla^2\mathbf{B})]\label{eq:emhd1},
\end{eqnarray}
where, $\mathbf{v}_e=-(\nabla\times\mathbf{B})/\mu_0n_0e$ is the electron fluid velocity. In addition to
ignoring the ion dynamics, Eq. (\ref{eq:emhd1}) assumes a uniform electron
number
density $n_0$ and the incompressibility of the electron fluid. Assuming $\omega << \omega_{pe}^2/\omega_{ce}$, displacement
currents are ignored.
In EMHD, the frozen-in condition of magnetic fluxes can break down only due to the
finite electron inertia (which is contained in the definition of $d_e \propto
\sqrt{m_e}$). In the absence of electron
inertia ($d_e\rightarrow 0$), Eq. (\ref{eq:emhd1}) represents the frozen-in condition of magnetic flux in an ideal electron flow. 

The equilibrium magnetic field is taken to be $\mathbf{B}_0 = B_{y0} \tanh(x/L)\hat{y} + B_{z0} \hat{z}$ corresponding to a current density $\mathbf{J}_0 = (B_{y0} /\mu_0 L)\,\mathrm{sech}^2 (x/L)\hat{z}$, where $L$ is the half thickness of the electron current sheet. For stationary ions, the electron fluid velocity is related to the current density by the relation $\mathbf{J} = -n_0 e\mathbf{v}_e$ . In the limit of cold electrons, the bipolar electrostatic electric
field co-located with the electron current sheet balances the Lorentz force in the current
sheet. Small deviations from charge neutrality in the electron current sheet can support the
bipolar electric field \cite{li2008} . This force balance is different from the force balance between pressure gradient and Lorentz force in the case of a Harris current sheet. Meanwhile, the bipolar electrostatic field and the force balance in electron current sheets have been observed in particle-in-cell simulations \cite{li2008,chen2011} , laboratory experiments \cite{yoo2013} , and space observations \cite{chen2009}.

Localized reconnection is initialized by adding a perturbation to all the equilibrium variables. Accordingly the
initial perturbation is assumed to have the form,
 
\begin{eqnarray}
\tilde{\psi}(x,y,z,t=0)&=&0.1\, e^{-(x^2+z^2)/2L^2}\sin(\pi y/l_y)\label{eq:inipert}
\end{eqnarray}
where $\tilde{\psi}$ denotes components of either $\tilde{\mathbf{B}}$ or $\tilde{\mathbf{v}}$. A perturbation (\ref{eq:inipert}) produces a single X-point in the reconnection (x-y) plane (z=0) being confined around z = 0 within a distance of $\sqrt{2}L$ in
the direction (z) perpendicular to the reconnection plane. 
The simulation box extends from $x = -l_x$ to $l_x$ , $y = -l_y$ to $l_y$ and $z = -l_z$ to $l_z$. The boundary conditions are periodic along y and z while the perturbations vanish at x boundaries far away from the central region of interest.

Simulations are carried out for a current sheet half thickness $L = d_e$ and guide fields from $B_{z0} = 0$ to $B_{z0} = 10 B_{y0}$ . The simulation box size $(2l_x \times 2l_y \times 2l_z)$ is
$10d_e \times 10d_e \times 160d_e$ with a grid resolution of 0.25 $d_e$ in each direction. The initial time step is $\omega_{ce}\Delta t = 0.01$. However, the time step can vary during the simulations in order to resolve the
largest velocity in the simulations according to the Courant condition with a Courant number=0.2.


Results are presented in normalized variables. The magnetic field is normalized by $B_{y0}$ ,
length by the electron inertial length $d_e$ , time by the inverse electron cyclotron frequency $\omega_{ce}^{-1} = (eB_{y0}/me)^{-1}$ , and the velocity by the electron Alfven velocity $v_{Ae} = d_e \omega_{ce}$ . Under this
normalization $\mathbf{J} = -\mathbf{v}_e$ holds.

\begin{figure}
\includegraphics[width=0.6\textwidth,height=0.7\textheight]{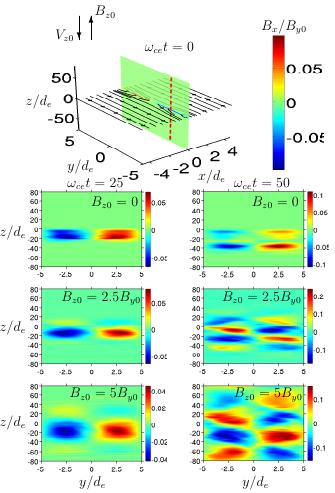}
\caption{\label{fig:bx3d} Normal component of magnetic field ($B_x$ ) in the plane x = 0 at $\omega_{ce}t=0$ (for all the simulations; top panel), $\omega_{ce} t = 25$ (left column) and $\omega_{ce} t = 50$  (right column) for $B_{y0} /B_{z0}$ = 0, 2.5 and 5. Projection of magnetic field lines in the plane z=0 and directions of equilibrium electron flow and guide field are shown in top panel only. The dashed line in top panel
is at $x=0$ and $y=-2.5 d_e$.}
\end{figure}

\begin{figure}
\includegraphics[width=0.8\textwidth,height=0.8\textheight]{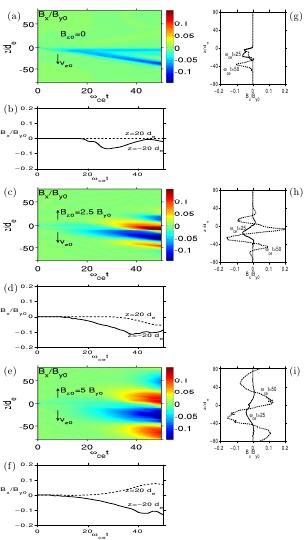}
\caption{\label{fig:bx_z_t} $B_x$ as a function of $z$ (along the dashed line,  $x=0, y=-2.5 d_e$, in top panel of Fig. 1) and time for $B_{z0}=0$ (a), $B_{z0}=2.5\,B_{y0}$ (c) and $B_{z0}=5\, B_{z0}$ (e). In (a), (c) and (e), the directions of equilibrium electron flow $V_{z0}$ and guide field $B_{z0}$ are indicated by arrows. Lineouts of $B_x$ along z-axis at $\omega_{ce}t=25$ and 50 in (g), (h) and (i).Lineouts of $B_x$ along time axis at $z=-20\, d_e$ and $20\, d_e$ in (b), (d) and (f).}
\end{figure}


\begin{figure}[t]
\includegraphics[width=0.75\textwidth,height=0.75\textheight]{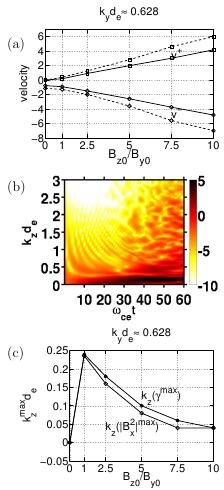}
\caption{\label{fig:v_sp}Speeds of spreading, $v^+$ (squares) and $v^-$ (circles) in the directions of the guide field and electron flow velocity respectively, as a function of the guide field strength (a). The spreading speeds are calculated  from both the nonlinear simulation (dashed line) and the linear theory (solid line). Evolution of the power in different $k_z$ for $B_{z0}=2.5\,B_{y0}$ (b). Variation with the guide field of the wave number $k_z$ for which linear growth rate is maximum (stars) and the power in $B_x$ is maximum (c).}  
\end{figure}

\begin{figure}
\includegraphics[width=0.75\textwidth,height=0.5\textheight]{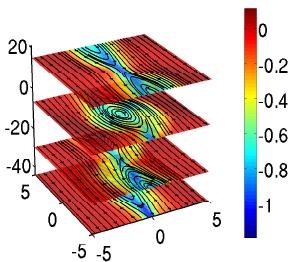}
\caption{Electron velocity $v_z$ (color) and projection of magnetic field lines (black lines) at $\omega_{ce}t=50$ in the planes $z/d_e$ = -6.5, -29.5, -44 and 15 for $B_{z0}=2.5 B_{y0}$ .
}
\end{figure}

\section{Spreading of localized magnetic reconnection\label{sec:spreading}}

In general, an electron current sheet is unstable to tearing and non-
tearing modes. For a finite guide field and $L = d_e$ , the growth rate of the fastest growing
non-tearing mode is larger but comparable to that of the 2-D tearing mode ($k_z = 0$) \cite{jain2015}. The power of the 
initial perturbation $\propto \sin(\pi y/l_y ) \exp(-z^2/2L^2 )$ added to the equilibrium peaks for $k_z = 0$ and $k_y d_e = 0.628$. This correspond to a most unstable 2-D tearing mode for $L = d_e$. Therefore the initial evolution of ECS is tearing-dominated causing magnetic reconnection. Later, non-tearing modes with finite $k_z$ will also grow and influence the ECS evolution.

Fig. \ref{fig:bx3d} shows the normal component of magnetic field ($B_x$ ), resulting from magnetic reconnection, in the plane $x = 0$ which is the mode-rational surface for the 2-D tearing mode ($k_z = 0$). Fig. \ref{fig:bx_z_t} again shows $B_x$ along the line $x=0, y=-2.5\,d_e$ as a function of $z$ and $t$. Reconnection first takes place in the region around $z = 0$ where the initial perturbation is localized. Then it spreads  along the guide field, away from z=0.  From Figs. \ref{fig:bx3d} and \ref{fig:bx_z_t}, the following features can immediately be obtained. (1) The spreading for zero guide field is different from that for non-zero guide field. For zero guide field, the spreading is unidirectional but becomes bidirectional as the strength of the guide field is increased. (2) For finite guide fields, reconnection spreads in a wave-like fashion with a speed and wavelength which increases with the strength of the guide field. (3) The spreading is asymmetric for small guide field and  reconnection spreads faster in the direction of the electron flow. 
On increasing the strength of the guide field, the spreading parallel and anti-parallel to the direction of the electron flow  increasingly becomes symmetric.

We calculate the speeds of the spreading, $v^+$ (along the guide field, $\hat{z}$) and $v^-$ (along the electron flow velocity, -$\hat{z}$) by noting down how far the reconnection signal has reached at a given time.  
Fig. \ref{fig:v_sp}a shows variations of $v^+$ and $v^-$ with the guide magnetic field. For a zero guide field, reconnection spreads only in the direction of the equilibrium electron flow with the peak equilibrium-flow-speed $v_{z0}=1$, ($v^+=0$ and $v^-=1$). For $B_{z0}=B_{y0}$, there is a small amount of spreading in the direction of the guide field but with a speed $v^+ < v_{z0}$ while the speed of spreading in the direction of equilibrium flow velocity $v^-$ is slightly  larger than $v_{z0}$. Further increasing the strength of the guide field up to $B_{z0}=10\,B_{y0}$, the speeds of the spreading in the two directions continue to increase reaching the values much larger than $v_{z0}$ and thus making the spreading almost symmetric. 

These features of the spreading of reconnection can be understood in terms of the propagation of the whistler and the flow induced wave modes. Electron flow induced wave modes can cause the propagation of the reconnection perturbation in the direction of the electron flow velocity with a speed which depends on the wave number of the perturbation \cite{huba2002,jain2013}. In the presence of a finite guide field, whistler wave modes propagate both parallel and anti-parallel to the guide field. The dispersion relation for two wave modes can  be obtained from EMHD local dispersion relation. The latter can be written as \cite{jain2015,jain04},
\begin{eqnarray}
\bar{\omega}&=&\frac
{k_z(d_e^2v''-v)\pm \sqrt{k_z^2(d_e^2v''-v)^2+4d_e^4\omega_{ce}^2(F''+k^2F)(F-d_e^2F'')/B_0^2}}
{2(1+k^2d_e^2)}
\label{eq:local_disp}
\end{eqnarray}
where $\bar{\omega}=\omega-k_zv$, $k^2=k_x^2+k_y^2+k_z^2$ and $F=\mathbf{k}.\mathbf{B}$. In the absence of the electron flow ($v=-B'$ and its higher derivatives vanish) Eq. (\ref{eq:local_disp}) becomes the well-known whistler-mode dispersion relation. 
 \begin{eqnarray}
\omega&=&\pm \frac
{k k_{||} d_e^2\omega_{ce}}
{1+k^2d_e^2}
\label{eq:whist_disp}
\end{eqnarray}

In the limit $\mathbf{B}\rightarrow 0$, one obtains the dispersion relation of  the flow induced wave modes \cite{das01}. 
 \begin{eqnarray}
\omega&=&\frac
{k_zd_e^2(v''+k^2v)}
{1+k^2d_e^2}
\label{eq:flow_disp}
\end{eqnarray}
Note that the phase velocity of the flow induced wave modes, Eq. \ref{eq:flow_disp}, is uni-directional (in the direction of the electron flow), while that of the whistler modes is bidirectional (parallel and anti-parallel to the mean magnetic field), Eq. (\ref{eq:whist_disp}). When both the guide field and electron flow are  present, one can expect that the spreading would have contributions from the two wave modes. In our simulations, the group velocities of the two wave modes will add up in the direction of  the electron flow while they will be subtracted in the direction of the guide field to provide the speed of the spreading. By this reasoning, the speeds of the spreading can be written as,
\begin{eqnarray}
v^+&=&v_w-v_f\,H(v_w-v_f)\label{eq:v+}\\
v^-&=&-(v_w+v_f)\label{eq:v-}
\end{eqnarray}
where $v_w$ and $v_f$ are the magnitudes of the group velocities of whistler and flow induced wave modes, respectively. The heaviside step function $H$ is unity when $v_w-v_f > 0$, or else it is zero.  

In order to verify Eqs. (\ref{eq:v+}) and (\ref{eq:v-}), we calculate $v_f$ and $v_w$ from linear theory of electron shear flow instabilities \cite{jain2015}. The calculations of $v_f$ and $v_w$ require wave numbers of the dominant modes in the simulations. The dominant modes in the simulations are the fastest growing mode and the modes in its neighborhood in the wave number space. These fast growing modes grow to attain maximum power. 
This is illustrated in Fig. \ref{fig:v_sp}b which shows evolution of the power in $k_z$ for $k_yd_e=0.628$ (initialized wave number along y-direction) and $B_{z0}=2.5\,B_{y0}$. In the late stage of the evolution, the power, initially peaked at  $k_z=0$, peaks around  $k_zd_e\approx 0.16$ which is close to the wave number of the fastest growing mode for $L=d_e$, $B_{z0}=2.5\, B_{y0}$ and $k_yd_e=0.628$. We show in Fig. \ref{fig:v_sp}c the variations of $k_z$ for which linear growth rate is maximum and for which power in $B_x$ peaks. The two wave numbers, one obtained from linear theory and the other from nonlinear simulations are in good agreement. For $B_{z0} > B_{y0}$, the value of $k_z$ drops. This is the reason why the wavelength of the wave spreading increases with the guide field.

Next we calculate $v_f=|d\omega_r/dk_z|$ ($\omega_r$ is the frequency of the fastest growing mode) and $v_{w}=|d\omega_w/dk_z|$ ($\omega_w$ is the whistler frequency) at $k_yd_e=0.628$ and values of dominant $k_z$ obtained from the simulations. The $v^+$ and $v^-$, obtained from Eqs. (\ref{eq:v+}) and (\ref{eq:v-}), are denoted by the squares and circle connected by dashed lines in Fig. \ref{fig:v_sp}a. The two speeds of the spreading are in agreement with the theoretical estimates for small guide field. The difference between the two estimates grows with the guide field. This could be due to the use of fixed $k_y$ and $k_z$ in linear estimates. The speed of spreading, however, can have contributions from other wave modes present in the simulation.

 
As the strength of the guide field is increased, the group velocity of whistler wave mode increases. The speeds of the spreading, $v^{+}$ and $v^-$, are given by the difference and sum of the group velocities of the flow induced and whistler wave modes, respectively. If the guide field is not large enough, so that the group velocities of the flow induced and whistler wave modes are comparable, the propagation and thus spreading is asymmetric being faster in the direction  parallel to the electron flow than in the anti-parallel direction. When guide field is large enough, the whistler group velocity dominates the net velocity. Thus, the spreading becomes increasingly symmetric and faster on increasing the strength of the guide field.


The wave-like spreading of reconnection causes the formation of X- and O-points in the z=constant
planes separated by half the wavelength of the reconnection wave propagation. This can be seen in Fig. 4 in which
magnetic field lines and electron flow velocity for $B_{z0} = 2.5 B_{y0}$ are shown at $\omega_{ce} t = 50$
in four planes $z/d_e =$ 15, -6.5, -29.5 and -44. These planes correspond to the z-locations
of four of the positive and negative peaks of $B_x$ at $\omega_{ce}t = 50$. The formation of X- and O-points is consistent with the sign of $B_x$ . From the top panel of Fig. 1, it can be seen that
negative sign of $B_x$ correspond to an X-point in the center of the x-y plane. In addition to
the formation of X- and O-points, the electron current sheet can be seen to undulate along
y. This is because other unstable non-tearing modes have grown to significant amplitude
by $\omega_{ce} t = 50$. Note that in earlier studies \cite{jain2013} , X- and O-points form alternately along the direction of electron flow when simulations are initialized with a non-localized perturbation $\propto \sin(\pi z/l_z )$ which has the wavelength equal to $2 l_z$ . In the present study, system chooses
to form alternate X- and O-points.



\section{Summary\label{sec:conclusion}}


We have shown that a patch of the localized magnetic reconnection spreads bi-directionally  in a wave like fashion when a finite guide field is present. The spreading is caused by the propagation of the flow induced and whistler wave modes away from the localized patch. For small guide fields, the spreading is asymmetric being faster in the direction of the electron flow. On increasing the guide field, the spreading becomes increasingly symmetric.
One can ask the question at what guide field the speed of spreading becomes completely symmetric, i.e., $|v^+|-|v^-|=0$. From Eqs. (\ref{eq:v+}) and (\ref{eq:v-}), the difference of the two speeds, $|v^+|-|v^-|=-2\,v_f$ (for $v_w>v_f$), is zero only when $v_f=0$. Thus there exist no guide field at which the two speeds are exactly equal. However, the spreading will be almost symmetric due to the dominance of the whistler phase speed over flow induced wave speed for very large guide field.  The wave-like spreading of reconnection causes the alternate formation of X- and O-points in the reconnection 
planes separated by half the wavelength of the reconnection wave.

The results presented here apply only to electron scales. Although collisionless magnetic reconnection initiates at electron scales, it is coupled to ion and even larger fluid scales. The spreading of reconnection will thus be influenced by the ion scale physics. Earlier simulation studies at the ion scales ignores electron physics and show the spreading of reconnection by non-dispersive Alfven waves for sufficiently large guide field \cite{shepherd2012}. Our studies show the spreading at electron scales by dispersive whistler waves whose phase velocity depends on the wave number. The speed of spreading in a physical situation, where both electron and ion dynamics are important, can be expected to be a hybrid of the group velocities of the Alfven and whistler waves. 


\begin{acknowledgments}
This work was supported by Max-Planck/Princeton Center for Plasma Physics at the Max Planck Institute for Solar System Research, Justus-von-Liebig-Weg-3, G\"ottingen, Germany and by the German Science Foundation CRC 963, project A03. 
\end{acknowledgments}

\appendix

\bibliography{references}
\end{document}